\documentclass[preprint,12pt]{elsarticle}

\usepackage{amssymb}
\usepackage{color}

\journal{Nuclear Instruments and Methods - A}

\begin{document}

\begin{frontmatter}



\title{General features of experiments on the dynamics of laser-driven electron-positron beams}


\author[qub]{J. R. Warwick }
\author[qub]{A. Alejo }
\author[qub]{T. Dzelzainis }
\author[slac]{W. Schumaker }
\author[qub]{D. Doria }
\author[luli]{L. Romagnani }
\author[impl]{K. Poder }
\author[impl]{J. M. Cole }
\author[qub]{M. Yeung }
\author[cuos]{K. Krushelnick}
\author[impl]{S. P. D. Mangles }
\author[impl]{Z. Najmudin }
\author[qub]{G. M. Samarin }
\author[stfc]{D. Symes }
\author[cuos,lanc]{A. G. R. Thomas }
\author[qub]{M .Borghesi }
\author[qub]{G. Sarri}
% author list taken from PRL minus sims guys

\address[qub]{School of Mathematics and Physics, Queen's University Belfast, University Road, Belfast, BT7 1NN, UK}
\address[slac]{SLAC National Accelerator Laboratory, Menlo Park, California 94025, USA}
\address[luli]{LULI, Ecole Polytechnique, CNRS, CEA, UPMC, 91128 Palaiseau, France}
\address[impl]{The John Adams Institute for Accelerator Science, Blackett Laboratory, Imperial College London, London SW72BZ, UK}
\address[cuos]{Center for Ultrafast Optical Science, University of Michigan, Ann Arbor, Michigan 481099-2099, USA}
\address[stfc]{Central Laser facility, Rutherford Appleton Laboratory, Didcot, Oxfordshire OX11 0QX, UK}
\address[lanc]{Lancaster University, Lancaster LA1 4YB, United Kingdom}

\begin{abstract}
The experimental study of the dynamics of neutral electron-positron beams is an emerging area of research, enabled by the recent results on the generation of this exotic state of matter in the laboratory. Electron-positron beams and plasmas are believed to play a major role in the dynamics of extreme astrophysical objects such as supermassive black holes and pulsars. For instance, they are believed to be the main constituents of a large number of astrophysical jets, and they have been proposed to significantly contribute to the emission of gamma-ray bursts and their afterglow. However, despite extensive numerical modelling and indirect astrophysical observations, a detailed experimental characterisation of the dynamics of these objects is still at its infancy. Here, we will report on some of the general features of experiments studying the dynamics of electron-positron beams in a fully laser-driven setup. 

\end{abstract}


\end{frontmatter}


\section{Introduction} % % % % % % % % % % % % % % % % % % % % % % % % % % % % %
\label{intro}

Pair-plasmas represent a unique state of matter, since they consist of negatively and positively charged particles
bearing the same mass and (absolute) charge. These objects have recently been gathering increasing interest in the academic community, not only for their unique properties (see for instance, Ref. \cite{Helander}) but also for the major role they play in the dynamics of a wide range of extreme astrophysical objects. For instance, magnetized electron-positron plasmas exist in pulsar magnetospheres \cite{magnetosphere}, in bipolar outflows in active galactic nuclei \cite{AGN}, at the center of our own galaxy \cite{galaxy}, and in the early universe \cite{early}.

Different schemes have been proposed to generate neutral pair plasmas in the laboratory. The first generation of a pair-plasma was achieved by Oohara and collaborators \cite{fullerene} by creating equal distributions of positively and negatively charged fullerene ions. Interestingly, this scheme allows for the long-term study of pair plasma dynamics, without having to deal with mutual annihilation, the main fundamental factor limiting the lifetime of an electron-positron plasma. More recently, active research is carried out in producing and confining low-energy populations of electrons and positrons. Whilst Penning traps can guarantee excellent confinement of either population, simultaneous confinement of both species beyond their Debye length cannot be achieved \cite{penning}. A solution to this problem has been identified in using toroidal magnetic fields either produced by levitated dipole configurations \cite{MIT} or stellarators \cite{stellarator}, as in the APEX project \cite{penning2}.

An alternative method for the generation of neutral electron-positron plasmas has been identified in exploiting the quantum cascade initiated by a laser-driven electron beam propagating through a solid target. Even though this method suffers from the difficulty of generating cold populations and confining them, promising results have been obtained in different configurations \cite{Chen1,Chen2,SarriPRL1,SarriNCOMM, SarriPPCF1,SarriPPCF2,SarriPRL2}, with, to-date, the first experimental demonstration of a neutral electron-positron beam able to show collective behaviour \cite{SarriNCOMM,SarriPRL2}. The details of the laser-driven generation of neutral electron-positron beams (EPBs) can be found elsewhere \cite{SarriPRL1,SarriNCOMM, SarriPPCF1,SarriPPCF2}; in this paper, we will instead focus our attention on some of the main experimental issues involved in detecting the dynamics of these objects, when they propagate through a background electron-ion plasma. 

The structure of the paper is as follows: in Section 2 we will discuss the general features of a typical experimental setup adopted in this class of experiments. In section 3, we will show how the laser-driven generation of EPBs simultaneously provides an efficient way of ionising the gas through which the EPB itself is to propagate. In Section 4 we will discuss 
the main concepts behind detecting EPB transverse instabilities using proton radiographic techniques while in Section 5 we will show some typical experimental results in this area. Finally in Section 6 we will discuss some limiting factors in directly detecting density modulations in a perturbed EPB while conclusive remarks will be presented in Section 7.

\section{Experimental Setup} \label{setup} % % % % % % % % % % % % % % % % % % %

\begin{figure}[!ht]
	\centering
	\includegraphics[width=\textwidth]{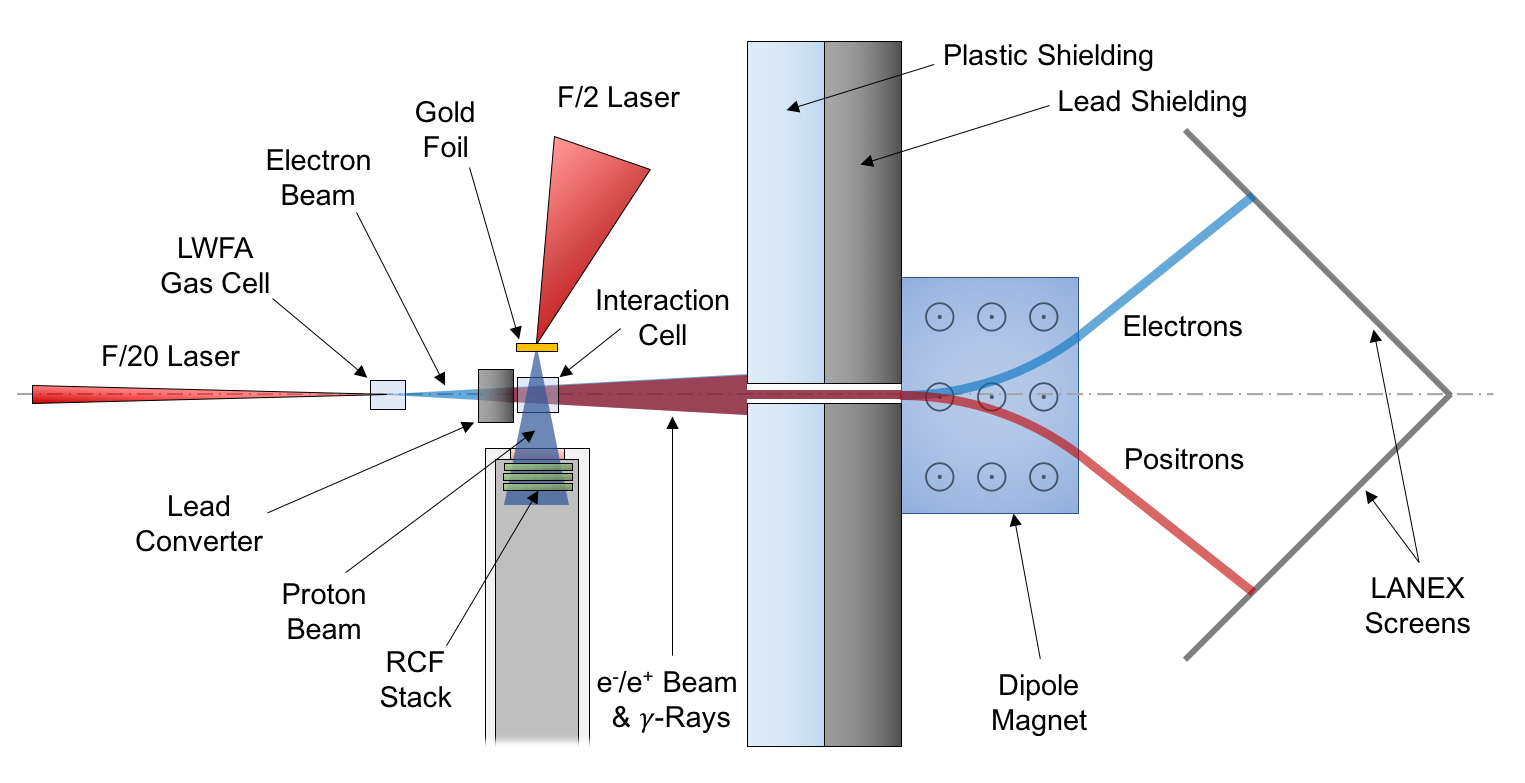}
	\caption{Typical experimental setup for the study of the dynamics of laser-driven EPBs}
	\label{fig:setup}
\end{figure}

A typical experimental setup designed to generate EPBs and study their dynamics is shown in Fig. \ref{fig:setup}. In its most general configuration, a high-power laser beam is loosely focussed at the edge of a low-density gaseous target in order to generate, via laser wakefield acceleration \cite{wakefield}, a short (beam duration of the order of tens of fs) and high energy (energy per particle in the region of hundreds of MeV) beam of electrons. A thick and high-Z solid target is then placed in the electron beam path in order to initiate an electromagnetic quantum cascade whose main by-products are electrons, positrons, and gamma-ray photons \cite{SarriPPCF1}. Numerical and experimental work has demonstrated that the percentage of positrons in the generated EPB can be seamlessly tuned from 0\% all the way up to 50\% by simply changing the thickness of the solid target \cite{SarriNCOMM}. Typical optimum parameters of the EPB at the source thus far include a source size of $D_0\simeq200-300$ $\mu$m, a duration in the range of $\tau\simeq$10s to 100s of fs, a number of leptons of the order of $N\simeq10^9$ - $10^{10}$, and an average divergence of $\theta\simeq30 - 50$ mrad. The broad spectrum of the EPB, virtually insensitive to the initial shape of the electron spectrum but mainly determined by the cascade itself, is well approximated by a Juttner-Synge distribution, with a typical average Lorentz factor of $\gamma\simeq 10-20$ \cite{SarriNCOMM,SarriPPCF1,SarriPRL2}. At the source, one can easily achieve a number density of the EPB of the order of $10^{16}$ cm$^{-3}$. However, due to the intrinsic energy-dependent divergence introduced by the cascade, this density drops quickly during the EPB propagation. Nonetheless, it is interesting to note that the possibility of exciting collective modes in the EPB is mostly dictated by the number of skin depths that are contained in it, either longitudinally or transversally. In the transverse direction, the number of skin depths ($\ell_S$) within the beam diameter is not influenced by the diameter ($D_\perp$) of the beam itself, since it can be expressed as: $D_\perp/\ell_S\approx 4.1\times10^{-4}\sqrt{N/(\gamma\tau\mbox{[fs]})}$ \cite{SarriNCOMM}, justifying the assumption that transverse collective behaviour is, in a first approximation, not influenced by the divergence of the beam. However, one must take into account that the broad spectrum of the EPB induces significant temporal spreading of the beam. This effectively is the main factor in determining the loss of transverse collective behaviour in the EPB as it propagates. 

Several theoretical works reported in the literature (see, as possible examples, Ref. \cite{SarriNCOMM,Silva,Bret,Dieckmann,Shukla}) have shown how the propagation of an EPB in a background electron-ion plasma might trigger the onset of a series of instabilities, the most important ones namely being the oblique, the two-stream, and the filamentation (sometimes loosely referred to as Weibel) instabilities. The competition between these instabilities is ruled by a series of beam and background plasma parameters, with the dominant one arguably being the ratio between the beam density $n_b$ and background plasma density $n_p$: $\alpha=n_b/n_p$. As a rule of thumb, oblique instabilities dominate for $\alpha\ll1$ whilst the filamentation instability dominates for $\alpha\geq1$ \cite{SarriNCOMM,Silva,Bret,Dieckmann,Shukla}. Two-stream instabilities are instead dominant for non-relativistic beams \cite{Bret} and will not be considered here.

The most direct measurables indicating the onset of instabilities are a spatial modulation of the electron and positron populations in the EPB, the generation of strong magnetic fields, and changes in the spectral shape of the EPB. However, the latter is to be expected only if the EPB is of sufficient length so that collective behaviour in the longitudinal direction can be triggered. This is not the case for experimental results reported in the literature \cite{SarriNCOMM,SarriPRL2} and it is generally not easy to achieve this condition with current laser systems. Even though it might be possible to achieve collective behaviour in the longitudinal direction in the near future, we will not discuss it here.
In the following sections, we will discuss the main experimental implications in measuring these quantities, after having discussed how the background gas gets effectively ionised after the interaction of the primary electron beam with the solid target.

\section{Ionisation of the background gas} % % % % % % % % % % % % % % %
\label{hyades}

\begin{figure}[!ht]
	\centering
	\includegraphics[width=1\textwidth]{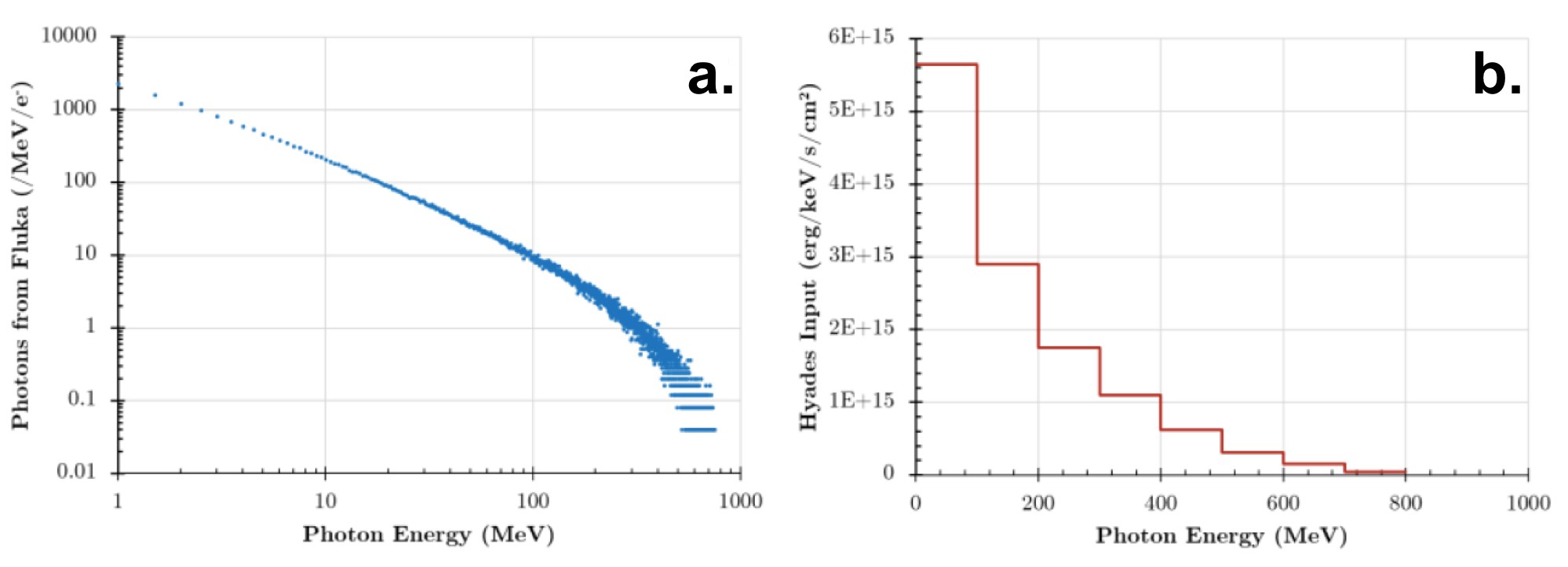}
	\caption{Details of the hydrodynamic simulation used to estimate the initial conditions of the gas in the second gas cell. (a) Gamma-ray spectrum at the exit of the Pb converter, as obtained from FLUKA simulations. (b) Input used in the Hyades simulation, after binning and conversion of the spectrum in (a) to suitable units (see text for details).}
	\label{fig:gamma}
\end{figure}

In order to experimentally observe EPB-driven plasma instabilities, it is crucial for the background gas within the second gas-cell to be fully ionised and form an ambient plasma, through which the EPB could interact. It must be noted here that, in addition to the EPB itself, the propagation of the ultra-relativistic electron beam through the solid target generates a bright and prompt flash of bremsstrahlung gamma-rays \cite{SarriPPCF1,SarriJPP,Schumaker}. This burst of gamma-ray photons photo-ionises the background gas; in the case of a He gas, and for typical experimental parameters, it is straightforward to achieve full ionisation, with typical electron temperatures of the order of 10s of eV. This conclusion is inferred from simulations performed using HYADES~\cite{HYADES}, a commercial 1-D hydrodynamics Lagrangian code frequently used to estimate the conditions of a plasma in laboratory astrophysical studies (for examples, see Refs.~\cite{ahmed,remington}).  As an example, we will discuss the experimental conditions presented in a recent publication \cite{SarriPRL2}. The simulation considers a $\gamma$-ray flash propagating through a 1~cm region filled with He gas at a backing pressure of 200~mbar.

\begin{figure}[!ht]
	\centering
	\includegraphics[width=1\textwidth]{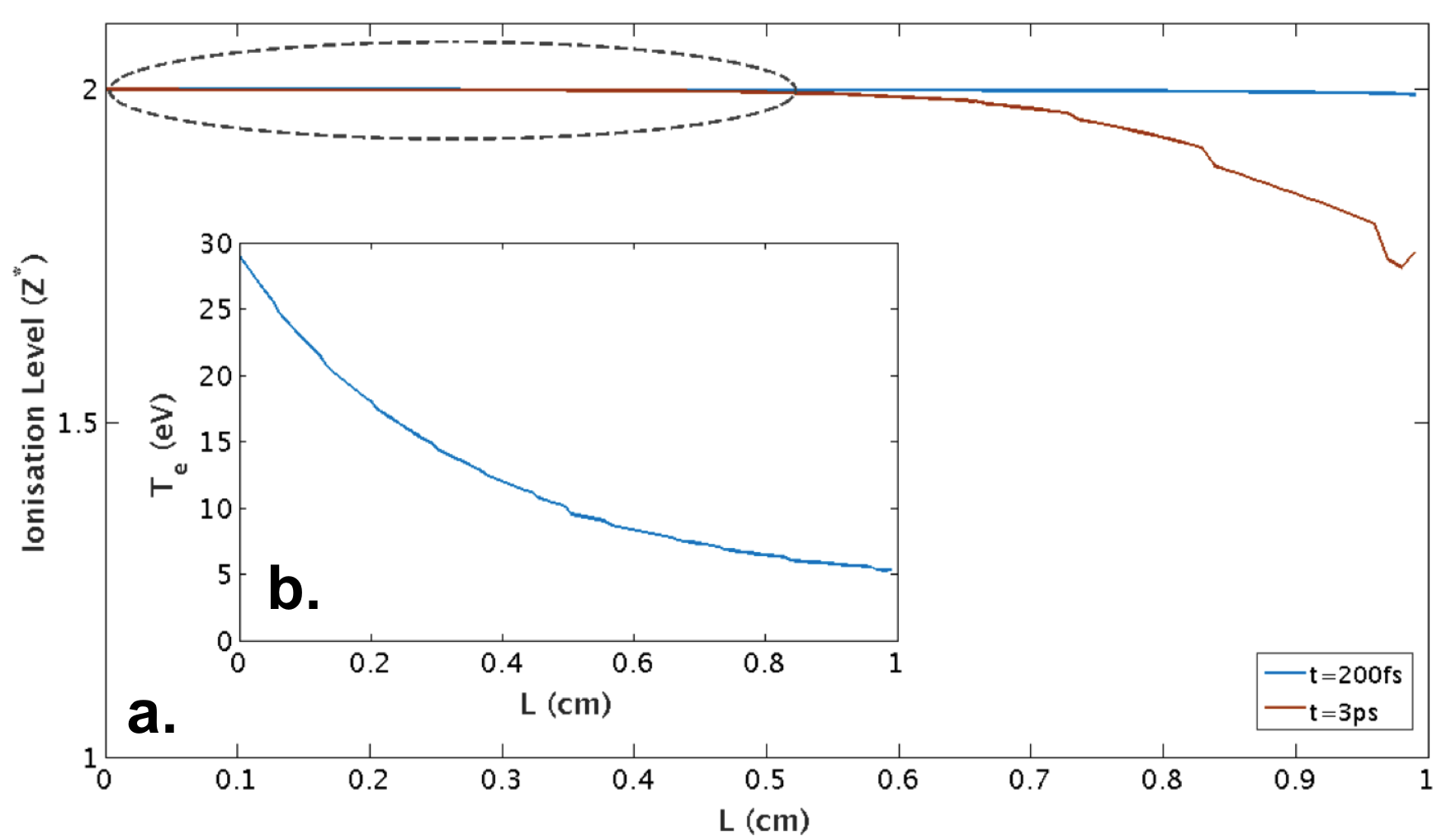}
	\caption{Results from hydrodynamic simulations on ionisation of the He gas in the second gas-cell, assuming a photon angular spectrum derived from FLUKA simulations. The ionisation level of the background gas as a function of length in the gas-cell is shown in (a), with the grey ellipse indicating the length over which the plasma is fully ionised.  Insert (b) shows the related electron temperature.}
	\label{fig:hyades}
\end{figure}

The spectrum of the $\gamma$-ray flash is obtained from FLUKA simulations~\cite{FLUKA}, considering the propagation of an electron beam analogous to those experimentally characterised, through 25~mm of lead (Fig.~\ref{fig:gamma}.a.).  The $\gamma$-spectrum is subsequently converted to a suitable format for the hydrodynamical simulation.  The $\gamma$-flash is modelled with a triangular temporal profile with a 100~fs FWHM duration.  Due to limitations in the number of energy ranges that could be simulated, the energy range is divided into 100~MeV bins (or \textit{photon group sources} in HYADES nomenclature).  The flux in each bin was calculated in units of erg/keV/s/cm$^2$, as required by the code, using the equation 
\begin{equation}
	F = \frac{E(\mathrm{erg})}{\Delta E(\mathrm{keV}) \tau(\mathrm{s})A(\mathrm{cm}^2)}
\end{equation}
where $E(\mathrm{erg})$ is the total energy of the photons in a bin (calculated from the spectrum obtained using FLUKA, $\frac{dN}{dE}$, as $\int_{E_0}^{E_1}E\frac{dN}{dE}dE$), $\Delta E$(keV) is the width of the energy bin in keV, $\tau$(s) is the duration of the gamma flash in s, and $A$(cm$^2$) is the beam size at the entrance of the gas cell in cm$^2$.  The final calculated input is shown in Fig.~\ref{fig:gamma}.b.

A period of 3~ps is simulated, with variable time steps internally calculated by the code taking into account the relevant time scales of the simulation.  The propagation of the gamma flash results in a prompt ionisation within the gas-cell with a plasma temperature of the order of 10 - 20 eV, as shown in Fig.~\ref{fig:hyades}

\section{Detection of magnetic fields} % % % % % % % % % % % % % % % % % % % % % % % % % %
\label{proRad}

Several numerical works reported in the literature indicate the onset of transverse filamentation in the beam, after propagating through a neutral background plasma. Without going in the details of the specific mechanisms involved (see, for instance, Ref. \cite{Shukla}, for details), it will suffice to note here that transverse modulations in the EPB result in localised current in the otherwise overall neutral beam. These beamlets of alternating current generate azimuthal fields that co-propagate with the EPB. Due to the fast time-scales involved - the EPB is only 10s to 100s of fs long and it is ultra-relativistic - it is indeed extremely challenging to detect these fields directly in the EPB; however, these fields can be transferred to the background plasma, where their detection is less problematic. A necessary condition for magnetisation of the background plasma is that the electron gyroradius be smaller than the typical scale of the magnetic field. In the EPB filamentation process, the magnetic field will grow with a spatial scale comparable to the skin depth of the beam whereas, in the non-relativistic limit, the electron gyroradius in the background plasma can be expressed as: $r_g = m_ev_{th}/eB$, where $v_{th}$ is the thermal velocity of the background electrons. The assumption of non-relativistic behaviour of the background plasma is justified by its relatively low temperature (tens of eV from HYADES simulations). The background plasma gets then magnetised if the magnetic field $B$ and the background plasma density $n_b$ obey the condition:
\begin{equation}
\xi=5\times10^{8}B\mbox{[T]}\sqrt{\frac{\gamma}{n_b\mbox{[cm}^{-3}\mbox{]}}}>1.
\end{equation}
For the parameters in Ref. \cite{SarriPRL2} we obtain $\xi\simeq 50$, fully justifying the assumption of the background plasma being magnetised. Once deposited in the background plasma, the magnetic field will decay either via collision-less processes (if the scale is comparable to the skin depth of the background plasma) \cite{Gruzinov} or collisionally (scales much larger than the skin depth of the background plasma). The latter situation effectively corresponds to $\alpha \ll 1$. In the case of collisional decay, the typical time scales involved would be proportional to the classical conductivity of the plasma \cite{SarriPRL2}; for typical experimental parameters, magnetic fields can then persist for hundreds of nanoseconds up to microseconds, a comfortable timescale to measure in current laser-plasma experimental setups. 

The spatial and temporal distribution of these fields can mainly be measured by looking at the change of polarisation in a transversely propagating optical probe (Faraday rotation) or with proton-based radiography arrangements. Without going into too much detail, it will suffice here to say that the rotation in polarisation induced by such fields embedded in a tenuous plasma will indeed be extremely small, whereas the proton deflections will be easily detectable. We will thus discuss here only the measurements of these fields via proton radiography techniques \cite{SarriNJP,borghesi}. As an additional note, it is worth mentioning that, on the typical timescales at which proton radiographies are collected, no electrostatic fields are to be expected, as discussed in Ref. \cite{SarriPRL2}.

\begin{figure}[!ht]
	\centering
	\includegraphics[width=0.7\textwidth]{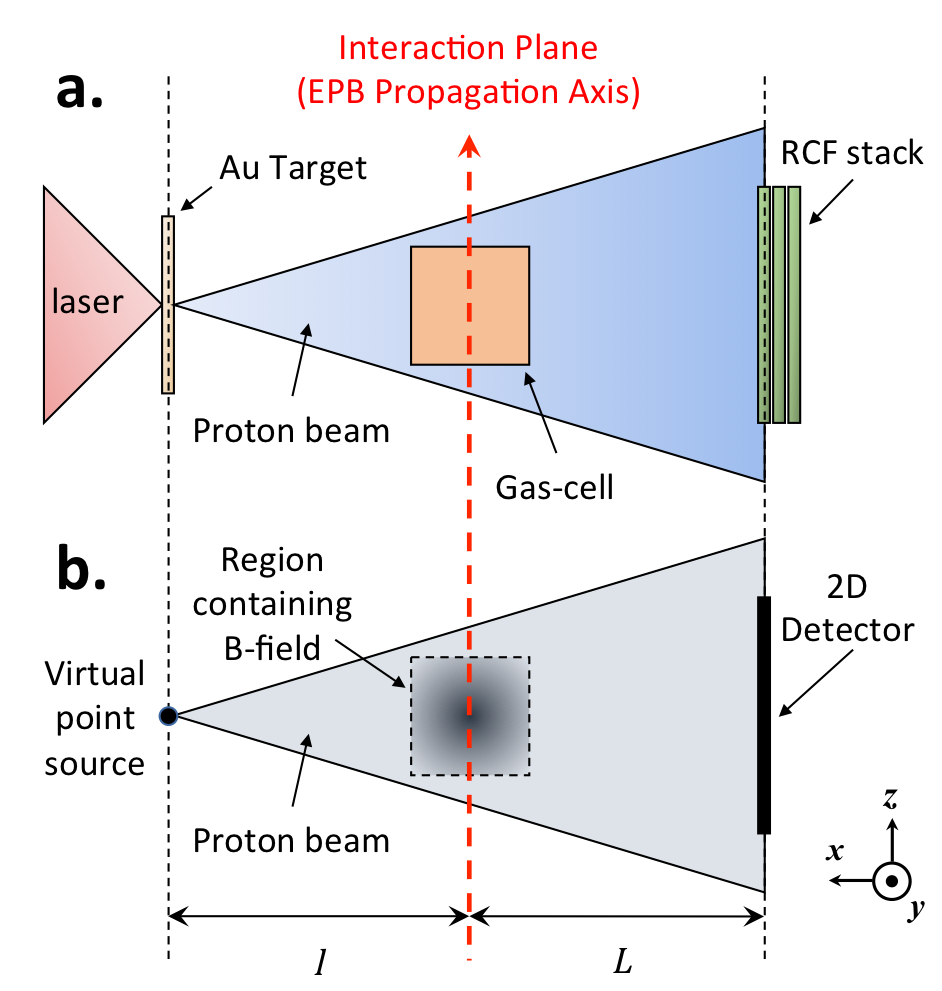}
	\caption{Sketch demonstrating how the experimental proton radiography setup (shown in a.) is simulated by the Particle Tracing code (shown in b.), with the same distances for source-to-interaction plane, $l$, and interaction plane-to-detector, $L$.  This allows for accurate 1:1 cross-comparison between the proton signal recorded on RCF and the 2-dimensional proton density pattern yielded from PT.}
	\label{fig:pt_diag}
\end{figure}

In the most general configuration for this technique (as also sketched in Fig. \ref{setup}), a high-power laser pulse is tightly focused upon a foil target of micron-scale thickness, producing a proton beam via Target Normal Sheath Acceleration \cite{RMP}.  These beams usually present a characteristic broad spectrum with a Maxwellian energy distribution, and a cut-off energy of the order of a few to tens of MeV.  A high-Z metallic target is usually preferred since it ensures a smoother spatial distribution of the proton beam. The configuration is set up to allow this probe beam to propagate through the interaction region, i.e. the plasma within the gas-cell.  Any electromagnetic fields present within this region deflect protons via the Lorentz force, leading to density fluctuations within the proton beam.  Beyond the interaction region, the protons impinge upon a multi-layer RCF detector stack, wherein the beam signal is recorded in the form of 2-dimensional proton density maps. The high degree of laminarity of the beam implies that the real source size of the protons, usually of the order of tens of microns, is equivalent to a much narrower virtual source size of the order of a few microns \cite{Borghesi2}. The typical spatial resolution of the diagnostic then results from the interplay between the size of the proton virtual source and the resolution of the proton detector. Moreover, the laminarity of the beam ensures a geometrical magnification of the phenomenon under interest of the detector \cite{SarriNJP}:
\begin{equation}
	 M \approx \frac{L+l}{l}
	\label{eqn:mag}
\end{equation}
where $l$ is the distance between the proton target and the interaction plane, and $L$ is the distance between the interaction plane and the detector (as shown in Fig.~\ref{fig:pt_diag}).  In the experiment reported in Ref. \cite{SarriPRL2}, $l = 8$~mm, and $L = 56$~mm, resulting in a magnification of $M \approx 8$. 

The broadband spectrum of the laser-driven protons implies that different spectral slices in the beam will traverse the region of interest at different times. The multi-layer arrangement of the RCF stack thus allows for a temporal multi-frame capability, even in a single shot. This is because different spectral bands will deposit most of their energy in the RCF layer that is in correspondence to the related Bragg peak. The main factors dictating the temporal resolution of each image are then: the proton pulse duration (usually in the range of a fraction of a picosecond up to a few picoseconds, depending on the parameters of the laser driving the protons) and the spread of energies deposited in each single RCF layer (usually of the order of $\pm$ 0.5 MeV). For low energy protons, the latter usually dominates, leading to temporal uncertainties from a few to tens of picoseconds.

In principle, this radiographic technique can be affected by a series of factors, namely: the shot-to-shot fluctuation in spectral and spatial properties of the proton beam and the presence of gas-filling through which the proton beam
must propagate. However, a change in spectral distribution is not important, since each RCF would be sensitive, in a first approximation, to the protons whose Bragg peak lies within the position of the RCF layer within
the stack \cite{SarriNJP}. Any subtle spectral effect on the radiographs is anyway taken into account on a shot-to-shot basis. Moreover, using a relatively thick and high-Z target for generating the proton beam guarantees a smooth spatial distribution in each spectral region. Finally, the presence of the gas-fill has a negligible effect on the properties of the proton beam.
In order to support this statement, we have performed simulations of the scattering induced by the gas-fill on the proton beam using the commercial Monte-Carlo scattering code SRIM \cite{SRIM}. These simulations provide an accurate estimate of the lateral straggling experienced by the protons due to scattering within the gas. Simulations indicate an induced broadening of the proton beam due to the presence of a 1cm thick gas-fill at a pressure of 200 mbar of approximately 2 micron (5 micron) for a 3.3 MeV (1.1 MeV) proton at the rear side of the gas-cell. This uncertainty is smaller than the intrinsic spatial resolution of the radiographs (of the order of 10 microns), and can thus neglected in the data analysis.

To diagnose the magnetic field distributions responsible for the observed proton deflections, a Particle Tracing (PT) code has to be used to simulate the proton radiography setup used in the experiment, as illustrated in Fig.~\ref{fig:pt_diag}. 

The code simulates the trajectories of a laminar beam of protons, from a virtual point source, through a 3-dimensional and time-dependent electromagnetic field distribution, to the detector plane.  The proton trajectories for a defined initial energy, $\varepsilon_p$,  are calculated by solving the non-relativistic equation of motion (Eqn.~\ref{eqn:eom}) 
\begin{equation}
	\frac{d\vec{v}_p}{dt} = \frac{e}{m_p}(\vec{E} + \vec{v}_p \times \vec{B})
	\label{eqn:eom}
\end{equation}
where $\vec{v}_p$ is the proton velocity, $e$ is the elementary charge, $m_p$ is proton mass, $\vec{E}$ is the electric field, and $\vec{B}$ is the magnetic field.  The PT code traces the proton propagation to the detector plane, where it produces a simulated 2-dimensional proton density map, that can be compared with the physical data. The 3-D electromagnetic field distribution, size, and strength within the interaction region can be altered and fine-tuned until a satisfactory match is achieved.

\section{Typical experimental results} % % % % % % % % % % % % % % % % % % % % % % % % %
\label{Results}

\begin{figure}[!ht]
	\centering
    \includegraphics[width=1\textwidth]{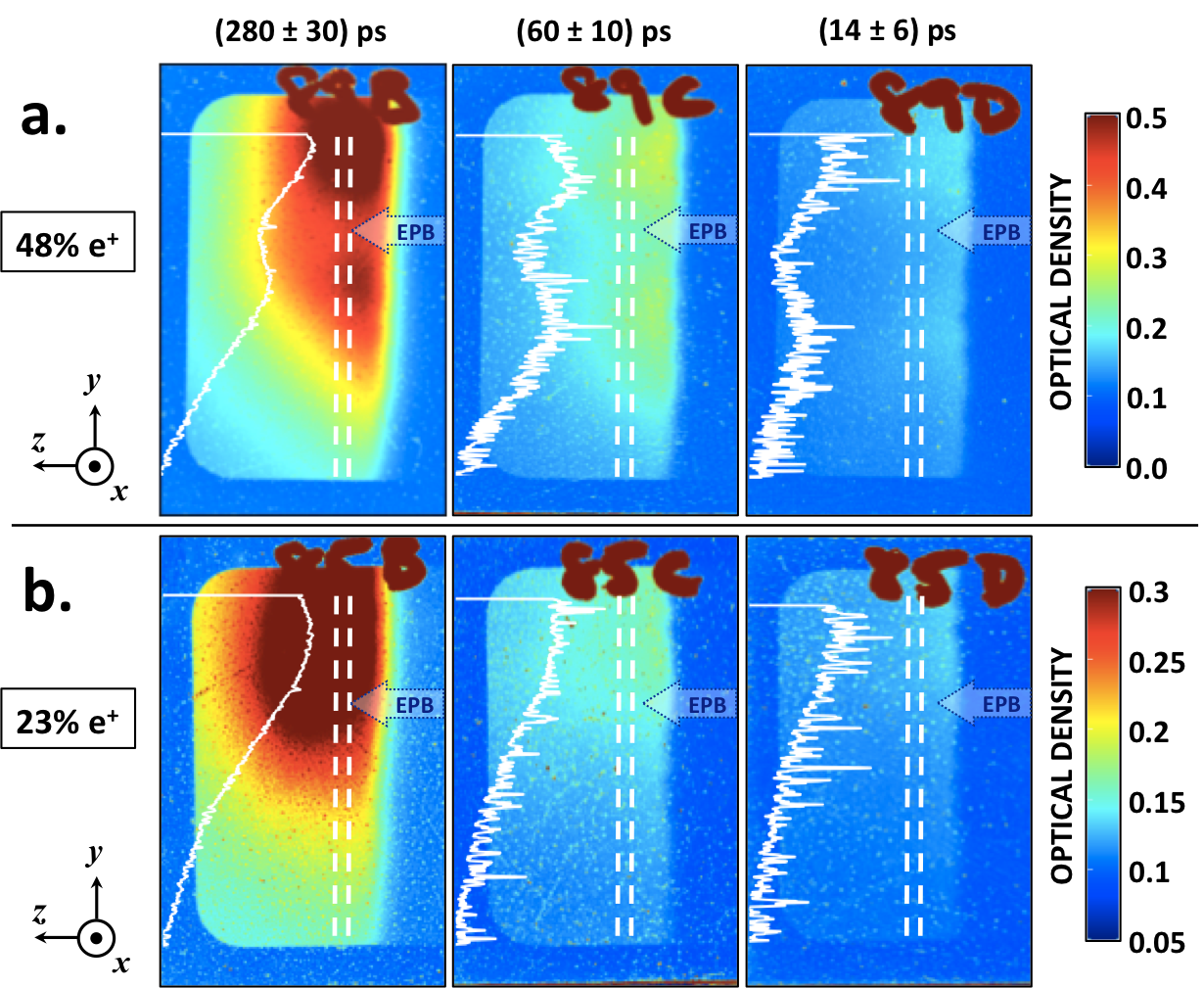}
	\caption{Proton radiographs of the background plasma after the propagation of a quasi-neutral EPB with 48\% e$^+$ population in row (a.), and a non-neutral EPB with 23\% e$^+$ population in row (b.).  Different columns correspond to different proton energies, and therefore, different probe times after the passage of the EPB (as labelled).  Each row corresponds to a single shot, highlighting the multi-frame capability of this radiographic technique.  The colour-bars represent the optical density on the RCFs for the respective rows, with a higher number corresponding to higher proton density).  Lineouts (white solid line) are taken from the regions between the white-dashed lines.  For an EPB with 23\% e$^+$ population (row b.), no clear modulation is seen on the proton radiographs (smooth monotonically decreasing profile, corresponding to the initial proton beam distribution). A pronounced modulation is observed for an EPB with 48\% e$^+$ population (row a.).  The modulation can be seen on all three layers of RCF indicating that the magnetic fields responsible for modulation are long-lived within the background plasma}
	\label{fig:rcf_compare}
\end{figure}

As an example of using the proton radiography technique discussed in the previous section in this class of experiments, we show in Fig. \ref{fig:rcf_compare} radiographs of the background plasma when a quasi-neutral (row a.) or a considerably non-neutral (frame b.) beam propagates through it. The data recorded via proton radiography shows that, as the EPB composition changed from a non-neutral beam to a quasi-neutral beam, the proton signal becomes progressively more perturbed, with a distinct observable modulation when quasi-neutrality is reached. Qualitatively, this is the first experimental indication that the generation and persistence of magnetic fields take place only for a quasi-neutral EPB.

\begin{figure}[!ht]
	\centering
	\includegraphics[width=\textwidth]{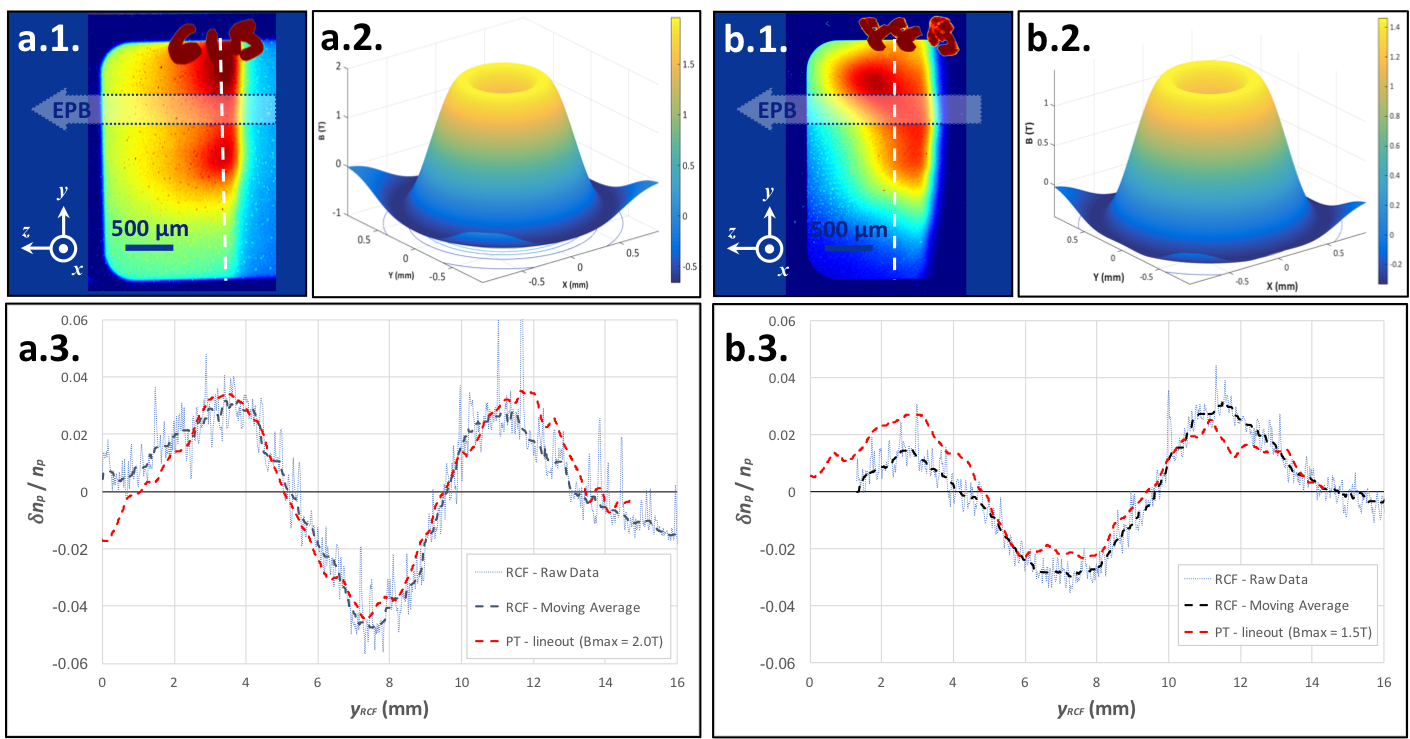}
	\caption{PT results for two separate shots taken with a quasi-neutral EPB (e$^+$ population of $48\pm5$\%).  Frames a.1. and b.1. show the raw experimental data converted to optical density, with the spatial scale referring to the interaction plane within the background plasma.  The EPB propagation direction is from right to left as indicated on the RCF.  The white dashed lines show the position at which lineouts where taken on the RCFs, and after background subtraction, are plotted in frames a.3. and b.3., both in raw format (blue dotted lines) and with moving average (black dashed lines) for easier reading.  Using the magnetic field distributions shown in frames a.2. and b.2. within the PT code, it was possible to extract lineouts (red dashed lines) that closely matched the experimental lineouts.}
	\label{fig:results}
\end{figure}

In order to extract quantitative information about the magnetic fields responsible for such proton perturbations, a particle tracing (PT) code was employed. The data presented in this article refer to the same experimental parameters reported previously \cite{SarriPRL2}.  The best match was found for a pure magnetic field with a distribution as shown in Fig.~\ref{fig:results}. (frames a.2. and b.2.), and modelled within the PT code by Eqn.~\ref{eqn:mag_dist}.
\begin{equation}
	B_\phi = B_{peak} \sin\left(\frac{2\pi\rho}{\rho_f}\right)\exp\left[-\left(\frac{\rho}{D_{beam}}\right)^4\right]
	\label{eqn:mag_dist}
\end{equation}

where $B_{peak}$ is the peak magnetic field, $\rho = \sqrt[]{x^2 + y^2}$ refers to the radial position in relation to the $z$-axis along which the EPB propagates (see Fig.~\ref{fig:results}), $\rho_f$ is the typical spatial scale of the magnetic field, and $D_{beam}$ is the diameter of the magnetic field distribution.  These magnetic field distributions are azimuthal around the EPB propagation axis, which is in the z-direction, from right to left on the RCFs, as shown in Fig.~\ref{fig:results}.

Frames a.1. and b.1. in Fig.~\ref{fig:results}. show the proton radiographs of the background plasma at a time of $280\,\pm\,30$~ps after a quasi-neutral EPB has passed through from right to left.  The white dashed lines indicate the locations where lineouts where taken.  The raw data is plotted in frames a.3. and b.3. (blue dotted line), with the moving average over plotted (black dashed line).  By adjusting the field parameters in Eqn.~\ref{eqn:mag_dist} it was possible to closely match lineouts garnered through the PT code with the experimental data (see red dashed lines in frames a.3. and b.3.).  For the radiograph presented in frame a., the magnetic field distribution was found to have a peak amplitude of $1.8\,\pm\,0.2$~T, with a spatial scale of $\rho_f\,=\,1.4\,\pm\,0.1$~mm, whilst the radiograph in frame b. was found to be best replicated by a magnetic field distribution with peak amplitude of $1.5\,\pm\,0.3$~T, with a spatial scale of $\rho_f\,=\,1.5\,\pm\,0.1$~mm. These values are consistent with those reported previously \cite{SarriPRL2}, confirming the robustness of the phenomenon and its detection, virtually unaffected by shot-to-shot fluctuations.

\section{Direct measurements of spatial density modulations in the EPB}

The process of transverse filamentation implies that the EPB will present strongly modulated electron and positron populations in its transverse profile. This intuitive statement is amply supported by Particle-In-Cell simulations of the interaction, as those reported in Ref. \cite{SarriNCOMM} for typical experimental parameters. One might then envisage the possibility of extracting the typical spatial features of the instability by directly measuring the transverse profile of the EPB after the propagation through the background plasma. However there are experimental factors that make this detection extremely difficult. First, one has to consider that the EPB co-travels with an intense burst of gamma-rays, with a photon number that greatly exceeds the number of particles in the EPB. Any scintillating material placed in the beam path will then predominantly interact with the gamma-ray beam, which will retain a smooth spatial distribution regardless of the dynamics of the EPB. This generates a considerably small contrast for any modulation in the EPB to be detected. One might think of separating the electron and positron populations with a magnetic dipole, and record them separately away from the main propagation axis of the gamma-ray beam. However, simple numerical calculations show that the external magnetic field necessary in order to obtain this separation (as used in the magnetic spectrometer in this class of experiments) is indeed sufficient to disrupt any small-scale modulation in the beam, restoring the transverse smoothness of the beam. Even though this difficulty is not of a fundamental nature and could be in principle overcome with a refinement of the experimental apparatus, there is a more subtle limitation that is tightly linked to the peculiar nature of a neutral matter-antimatter beam. It is relatively straightforward to see, from Particle-In-Cell simulations, that the electron and positron filaments are completely symmetrical, so that regions with a null number of electrons are exactly overlapped with a positron filament, and vice-versa. This should not be surprising, given the exact same mobility of the negatively and positively charged constituents of the EPB. Given that, in the ultra-relativistic case, typical particle detectors are virtually insensitive to the sign of the charge of the particle impinging on it ($\leq2$\% difference \cite{Carson}), they would then not be able to display non-uniformities in the charge distribution of the beam, but only show the number density, which is kept constant even in the case of filamentation. From these simple considerations, one can then argue that the most suitable measurable to indicate onset of transverse instabilities in the EPB is the magnetic field left in the background plasma.    

\section{Conclusions} % % % % % % % % % % % % % % % % % % % % % % % % % % % % % %
\label{conc}

In this article, we have discussed some of the key experimental issues concerning the detection of instabilities experienced by a laser-driven electron-positron beam as it propagates through a background electron-ion plasma, in conditions of relevance to the dynamics of pair-dominated astrophysical scenarios such as astrophysical jets and pulsar atmospheres. 
According to recently published experimental results in this area, we identify the magnetic fields left in the background plasma as the most suitable indicator of the onset of transverse instabilities and we have discussed some of the main characteristics of the detection of these fields via proton radiography. 

\section{Acknowledgments} % % % % % % % % % % % % % % % % % % % % % % % % % % % % % %
\label{ack}
The authors wish to acknowledge support from EPSRC (grants: EP/N022696/1, EP/N027175/1,
EP/L013975/1, EP/N002644/1,and EP/P010059/1). The authors would also like to thank the staff at CLF for their technical support.

\end{document}